\documentclass[11pt,a4paper]{article}
\usepackage{amsmath,amssymb,hyperref,epsfig,multicol}
\newcommand{\alg}[1]{\mathfrak{#1}}

\newcommand{\beq}{\begin{equation}}
\newcommand{\eeq}{\end{equation}}
\newcommand{\bea}{\begin{eqnarray}}
\newcommand{\eea}{\end{eqnarray}}

\newcommand{\be}{\begin{equation}}
\newcommand{\ee}{\end{equation}}

\usepackage{epsfig,multicol}
\begin{document}

\begin{titlepage}
\begin{flushright} 
MIT-CTP 3809\\
\end{flushright}
\mbox{ }  \hfill 
\vspace{5ex}
\Large
\begin {center}     
{\bf Classical $r$-matrix of the $\alg{su(2|2)}$ SYM spin-chain}
\end {center}
\large
\vspace{1ex}
\begin{center}
Alessandro Torrielli
\end{center}
\vspace{1ex}
\begin{center}
Center for Theoretical Physics\\
Laboratory for Nuclear Sciences\\
and\\
Department of Physics\\
Massachusetts Institute of Technology\\
Cambridge, Massachusetts 02139, USA\\ 
\vspace{1ex}
\texttt{torriell@mit.edu}

\end{center}
\vspace{4ex}
\rm
\begin{center}
{\bf Abstract}
\end{center} 
\normalsize 
In this note we straightforwardly derive and 
make use of the quantum $R$-matrix for the $\alg{su(2|2)}$ 
SYM spin-chain in the manifest $\alg{su(1|2)}$-invariant formulation, 
which solves the standard quantum Yang-Baxter equation,
in order to obtain the correspondent (undressed) 
classical $r$-matrix from the first order 
expansion in the ``deformation'' parameter $2 \pi / \sqrt{\lambda}$,
and check that this last 
solves the standard classical Yang-Baxter equation. We analyze its bialgebra 
structure, its
dependence on the spectral parameters and its pole structure. 
We notice that it still preserves an $\alg{su(1|2)}$
subalgebra, thereby admitting an expression in terms of a combination of
projectors, which  
spans only a subspace of $\alg{su(1|2)} \otimes \alg{su(1|2)}$. We study 
the residue at its simple pole at the origin, and comment on the applicability 
of the classical Belavin-Drinfeld type of
analysis. 

\vfill
\end{titlepage} 

\section{Introduction}

Integrability in AdS-CFT \cite{MZ} 
has revealed itself as one of the most promising 
developments towards a proof of the conjecture. Among the preminent results
is the derivation of a scattering matrix \cite{Beisert} whose tensorial 
structure is fixed by centrally extended $\alg{su(2|2)} \oplus \alg{su(2|2)}$
symmetry (see also \cite{nlin}), and whose dressing factor \cite{dress} is 
constrained by the Hopf-algebraic analog of what crossing symmetry is 
for relativistic systems \cite{Janik}. 
Remarkable advances in the exact determination of this phase factor
have been recently made \cite{AF,transcend}, such an impressive agreement
seemingly supporting the idea of a powerful algebraic structure 
underlying the full planar integrability of the model 
(see \cite{Yangian} and the 
recent 
\cite{malda,FZ,qpoincare}).
 
The presence of a Hopf algebra structure \cite{spanish,noi}
suggests that the full $S$-matrix 
might be a representation of the Universal $R$-matrix for a yet to be 
discovered bialgebra, which such an $R$-matrix would endow with a 
triangular structure. In order to make progress in understanding this 
construction, it is normally useful to study the classical limit of
the $R$-matrix, in terms of a deformation around the identity. Under 
certain assumptions (among which some non-degeneracy condition), 
powerful theorems allow a complete 
classification of solutions of the classical Yang-Baxter equation \cite{Drin}.
In particular, Belavin and Drinfeld \cite{Drin} considered solutions
$r(u_1,u_2)$
to the CYBE assuming values in $\alg{g} \otimes \alg{g}$, with $\alg{g}$ 
a finite-dimensional simple
Lie algebra, which are of difference form, namely they depend only on the 
difference $u$ of the two spectral parameters, $u = u_1 - u_2$. 
They introduced the additional hypothesis of
non-degeneracy, i.e. one of the three equivalent conditions: {\it (i)} 
the determinant of the matrix formed by the coordinates of the tensor $r$ 
is not identically zero, {\it (ii)} $r$ has at least one pole in the 
complex variable $u$, and it does not exist a Lie subalgebra $\alg{g'}$
such that $r$ is an element of 
$\alg{g'} \otimes \alg{g'}$ for any $u$, or {\it (iii)} $r$ has a 
first order pole in $u=0$, with residue of the form 
$c \sum_\mu I_\mu \otimes I_\mu$
with $c$ a complex number and $I_\mu$ a basis in $\alg{g}$ orthonormal with 
respect to a chosen nondegenerate invariant bilinear form. 
Such a residue can be identified with the quadratic Casimir operator 
in the tensor product of two copies of the Lie algebra.
Under these 
requirements, they proved that such solutions satisfy the unitarity 
condition $r_{12}(u) = - r_{21} (- u)$, and extends 
meromorphically to the entire $u$ complex plane.
All the poles of $r(u)$ are simple, and form a lattice $\Gamma$ 
in the complex plane.
Furthermore, modulo automorphisms, one has three possible types of solutions:
elliptic (if $\Gamma$ is a two-dimensional lattice), 
trigonometric (if $\Gamma$ is one-dimensional), or rational functions
(if $\Gamma = 0$). From the knowledge of such a classical $r$-matrix, 
there is a standard procedure to construct an associate
Lie bialgebra, in terms of so-called Manin triples (see for example 
\cite{Etingof} and references therein). This has to play the role 
of the enlarged symmetry algebra one is looking for. Work on the 
extension of these results to the more complicated case of simple Lie 
superalgebras began shortly after \cite{serg}
(see also \cite{zgb}).

The lack of difference form of the AdS-CFT $S$-matrix represents a source of
rich structure, which evades Belavin and Drinfeld's assumptions, and
its classification appears like an open problem\footnote{One can see 
\cite{nonadd} for work on $R$-matrices dependent on 
non-additive parameters.}.
In this note, we would like to set the tools for such an approach. 
We first obtain a form
of the quantum $R$-matrix in the manifest $\alg{su(1|2)}$-invariant
formulation, by performing in each entries of 
Beisert's spin-chain 
$S$-matrix \cite{Beisert} a transformation from excitation states 
$|\phi^1\rangle$,$|\phi^2\rangle$ to states $|\phi\rangle$, $|\chi\rangle$ 
\cite{Beisert,Janik} in initial and final states, 
producing the appearance of additional momentum-dependent phase 
factors upon consistent reabsorption of all the $Z$-markers. 
What we are left with is a genuine solution of the standard 
quantum Yang-Baxter equation (QYBE), in the spirit of the canonical $R$-matrix
of 
\cite{FZ}
to which this one should be related via a non-local basis 
transformation\footnote{We 
thank G. Arutyunov and S. Frolov for email exchange about this point.}.
The choice of the
$\alg{su(1|2)}$ formulation allows us to make direct contact with the 
Hopf-algebraic construction performed in \cite{noi}. Furthermore, it leaves a 
large number of generators undeformed, 
and singles out a 
decomposition by means of projectors, in terms of which it is quite simple to
perform the
consequent analysis of the classical $r$-matrix. The outcome is a formalism 
where the {\it dynamic} \cite{Beisert}
nature of the spin chain has been completely translated into
algebraic properties, and the presence of the lenght-changing operators 
has been totally removed, their action being fully
implemented into the resulting bialgebra.

Then, we take the strong coupling regime of the spin-chain, and expand 
the solution as

\begin{equation}
R_\alg{su(1|2)} \sim 1 + \zeta r_\alg{su(1|2)},
\end{equation}
which allows us to 
extract the classical $r$-matrix $r_\alg{su(1|2)}$ (the subscript stands 
for the $\alg{su(1|2)}$-basis), our deformation parameter
being 
$\zeta = 2 \pi / \sqrt{\lambda}$. In string theory language, the 
notion of
classical $r$-matrix would correspond in physical terms, taking into account
the contribution 
coming from the dressing factor, 
to the tree-level string sigma-model scattering matrix. This was computed 
(in a natural string basis) in
\cite{Zaretal}. Here, in more mathematical terms, 
we really try to identify $\zeta$ as the expansion parameter leading to 
a Poisson structure deformation (quantization parameter)\footnote{We thank 
M. Zabzine for discussions 
about this point.} \cite{Etingof}.     

We make use of 
the parametrization in \cite{AF} and keep fixed their $x$-parameter, which 
becomes the spectral parameter for the classical $r$-matrix. 
We check that this $r_\alg{su(1|2)}$ satisfies the standard
classical Yang-Baxter equation (CYBE). 
We stress that, being interested in symmetries of the classical $r$-matrix, 
we neglect in all computations the dressing factor, which 
amounts to adding to the classical 
$r$-matrix
a term proportional to the identity, thereby dropping from the CYBE. 

We then derive the first order coproduct relation 
from expansion of the full Hopf-algebraic one \cite{noi}.
$r_\alg{su(1|2)}$ still preserves an undeformed $\alg{su(1|2)}$ 
subalgebra, thereby admitting an expansion in terms of projectors onto 
irreducible representations of $\alg{su(1|2)} \otimes \alg{su(1|2)}$. 
We make use of a simpler Casimir operator with respect to \cite{Janik}.
One of the projectors has zero coefficient, therefore the $r$-matrix projects 
onto a subspace of $\alg{su(1|2)} \otimes \alg{su(1|2)}$.

We then examine the poles of the classical $r$-matrix: apart from 
uninteresting 
singularities in the parametrization, there is one pole for coincident 
spectral parameters, 
at which we calculate the residue. After a change of variables, the residue 
assumes the form of the Casimir of the algebra $\alg{gl(2|2)}$ shifted by
the identity. The appearance of this non-simple Lie algebra
naively seems to put 
into troubles the applicability of a Belavin-Drinfeld type of analysis.
This, in turn, might have been expected by considering the important role 
the central extensions have to play in the structure of the resulting Hopf 
algebra \cite{spanish,noi,qpoincare,Fabian}. Still, the presence of a Casimir 
operator indicates it might be possible to adapt the Manin triple
procedure \cite{Etingof} 
to the present case. Very recent developments point in fact towards 
the direction of a particular Yangian symmetry underlying the problem
\cite{BeisYang}.

Another pole is found when the 
two eigenvalues corresponding to the two out of three surviving projectors
coincide. The study of the analytic singularities 
of the full
quantum $S$-matrix, and in general of the dynamics ensuing from it, 
has been the subject of an intense work \cite{Dorey}.
We conclude with some comments, and appendices where we collect the bulk of 
the formulas.
 
\section{Quantum $R$-matrix $R_\alg{su(1|2)}$}
The centrally-extended $\alg{su(2|2)}$ commutation relations are as follows
\cite{Beisert}:

\begin{eqnarray}
[\alg{R}^a_b, \alg{J}^c] &=& \delta^c_b \alg{J}^a - \frac{1}{2} \delta^a_b \alg{J}^c,\nonumber\\
\relax
[ {\alg{L}}^\alpha_\beta, \alg{J}^\gamma ] &=& \delta^\gamma_\beta \alg{J}^\alpha - \frac{1}{2} \delta^\alpha_\beta 
\alg{J}^\gamma,
\end{eqnarray}
\begin{eqnarray}
\label{algebra}
\{\alg{Q}^\alpha_a,\alg{S}^b_\beta\} &=& \delta^b_a \alg{L}^\alpha_\beta + \delta^\alpha_\beta \alg{R}^b_a +  
\delta^b_a  \delta^\alpha_\beta \alg{C}, \nonumber\\
\{\alg{Q}^\alpha_a,\alg{Q}^\beta_b\} &=& \epsilon^{\alpha\beta}\epsilon_{ab}\alg{P}, \nonumber\\
\{\alg{S}^a_\alpha,\alg{S}_\beta^b\} &=& \epsilon_{\alpha\beta}\epsilon^{ab}\alg{K}\, .
\end{eqnarray}

The ($2|2$) 4-dimensional representation obtained by Beisert 
in \cite{Beisert} is labelled by 4 parameters $\text{a}$, $\text{b}$, $\text{c}$, $\text{d}$ with the constraint $\text{a}\text{d}-\text{b}\text{c}=1$. A basis of the representation space is provided by the vectors $|\phi^a\rangle$ (even) and $|\psi^\alpha\rangle$ (odd), with $a=1,2$ and $\alpha=1,2$, and the generators' action is 

\begin{eqnarray}
&&\alg{R}^a_b |\phi^c\rangle = \delta^c_b |\phi^a\rangle - (1/2) \delta^a_b |\phi^c\rangle,\nonumber\\
&&\alg{L}^\alpha_\beta |\phi^\gamma\rangle = \delta^\gamma_\beta |\phi^\alpha\rangle - (1/2) \delta^\alpha_\beta |\phi^\gamma\rangle
\end{eqnarray}
for the even ones, and
\begin{eqnarray} 
&&\alg{Q}^\alpha_a |\phi^b\rangle = \text{a} \, \delta^b_a |\psi^\alpha\rangle,\nonumber\\
&&\alg{Q}^\alpha_a |\psi^\beta\rangle = \text{b} \, \epsilon^{\alpha \beta} \epsilon_{a b} |\phi^b Z^+\rangle,\nonumber\\
&&\alg{S}^a_\alpha |\phi^b\rangle = \text{c} \, \epsilon^{a b} \epsilon_{\alpha \beta} |\psi^\beta Z^-\rangle,\nonumber\\
&&\alg{S}^a_\alpha|\psi^\beta\rangle = \text{d} \, \delta^\beta_\alpha |\phi^a\rangle
\end{eqnarray}
for the odd ones. We reorganize the indices $a,\alpha$ in a unique index
$A=1,2,3,4$ such that $a=1$ corresponds to $A=1$, $a=2$ to $A=2$, $\alpha=1$ to $A=3$ and $\alpha=2$ to $A=4$. We will adopt the $\alg{su(1|2)}$-formulation
\cite{Beisert,Janik}, in which the generators' action is redefined in such a way that no $Z$-markers appear, instead their effect is taken into account by a nontrivial coproduct (see \cite{noi} for details). 
The Hopf algebra structure is obtained by singling out an 
$\alg{su(1|2)}$ sub-algebra with generators $\alg{J} \in \{\alg{R}^1_1,{\alg{L}}^\alpha_\beta,
\alg{Q}^1_1, \alg{Q}^2_1, \alg{S}^1_1, \alg{S}^1_2, \alg{C}\}$. The coproduct on these generators is trivial, 

\begin{equation}
\label{coprod1}
\Delta(\alg{J}) = \alg{J} \otimes 1 + 1 \otimes \alg{J}.
\end{equation} 
The remaining generators split 
according to $\alg{D}_+ \in \{\alg{S}^2_1, \alg{S}^2_2, \alg{R}^2_1, \alg{K}\}$ and $\alg{D}_- \in \{\alg{Q}^1_2, \alg{Q}^2_2, \alg{R}^1_2, \alg{P}\}$, then one has 

\begin{eqnarray}
\label{coprod2}
&&\Delta(\alg{D}_+) = \alg{D}_+ \otimes e^{ip} + 1 \otimes \alg{D}_+,\nonumber\\
&&\Delta(\alg{D}_-) = \alg{D}_- \otimes e^{-ip} + 1 \otimes \alg{D}_-.
\end{eqnarray}
One identifies $e^{ip} = 1 + (\alg{K}/\beta)$ and $e^{-ip} = 1 + (\alg{P}/\alpha)$ as a physical constraint \cite{Beisert}. Up to a rescaling of state vectors this reads $\text{a}=1$, $\text{b} = - \alpha (1 - \frac{x^-}{x^+})$, $\text{c} = \frac{i \beta}{x^-}$, $\text{d} = - i (x^+ - x^-)$, $x^+ + \frac{\alpha \beta}{x^+} - x^- - \frac{\alpha \beta}{x^-} = i$, $\alpha \beta = \frac{g^2}{2}$. In this way the coproduct is symmetric on the center. Antipode and counite are specified in \cite{noi}.   

The $R$-matrix intertwining this coproduct, namely solving the equation

\begin{equation}
\label{coprrel}
\Delta^{op}(x) \, R = P \Delta(x) R = R \Delta(x)
\end{equation}
($P$ being the graded permutation operator) 
for any generator $x$ of the centrally extended $\alg{su(2|2)}$ algebra,  
and satisfying the standard quantum Yang-Baxter equation, is easily obtained from Beisert's $S$-matrix as $R = P S$, and by decorating it with suitable phase factors which come from re-expressing initial and final $|\phi^2\rangle$ states as $|\chi\rangle = |\phi^2 Z^+\rangle$ states, in the spirit of the $\alg{su(1|2)}$ formulation. We will call this R-matrix  
$R_\alg{su(1|2)}$, $R_\alg{su(1|2)} |v_1^A\rangle \otimes |v_2^B\rangle = \sum_{C,D=1}^4 R^{AB}_{CD} |v_1^C\rangle \otimes |v_2^D\rangle$. Pedices $1,2$ correspond to chosen representations. The entries of $R_\alg{su(1|2)}$ are reported in the Appendix. We stress the fact that we neglect everywhere in this paper the dressing factor, which has no effect on the quantum (and the classical) Yang-Baxter equation.   

Since there is an $\alg{su(1|2)}$ subalgebra with undeformed coproduct, one can express this very same $R$-matrix as $R_\alg{su(1|2)}= \sum_{i=1}^3 S_i P_i$, where $P_i$ are the projectors onto the irreducible representations of $\alg{su(1|2)} \otimes \alg{su(1|2)}$. In order to construct such projectors, we use a simpler Casimir with respect to the one used by Janik \cite{Janik}, namely we take

\begin{eqnarray}
C_{12} = \frac{1}{2} (\alg{Q}^1_1 \otimes \alg{S}^1_1 + \alg{Q}^2_1 \otimes \alg{S}^1_2 - \alg{S}^1_1 \otimes \alg{Q}^1_1 - \alg{S}^1_2 \otimes \alg{Q}^2_1) + (\alg{R}^1_1 + \alg{C}) \otimes (\alg{R}^1_1 + \alg{C}) - \frac{1}{2} \alg{L}^\alpha_\beta \otimes  \alg{L}^\beta_\alpha,
\end{eqnarray}   
whose eigenvalues are\footnote{In \cite{Janik} the Casimir is constructed 
multiplying the (trivial) coproduct of the $\alg{su(1|2)}$ generators. We 
thank the author for clarifications on this point.}
 
\begin{eqnarray}
\label{proj}
&&\lambda_1 = (1 + \text{b}_1 \text{c}_1)(1 + \text{b}_2 \text{c}_2),\nonumber\\
&&\lambda_2 = \frac{\text{b}_1 \text{c}_1 + \text{b}_2 \text{c}_2}{2} + \text{b}_1 \text{c}_1 \text{b}_2 \text{c}_2,\nonumber\\
&&\lambda_3 = \text{b}_1 \text{c}_1 \text{b}_2 \text{c}_2.
\end{eqnarray}
This Casimir is related to the one used by Janik as $C_{Janik} = 2 C_{12} + ({C^{(2)}}_1 + {C^{(2)}}_2 ) 1\otimes1$, if ${C^{(2)}}_i$ is the quadratic Casimir of $\alg{su(1|2)}$ in representation $i$. Selected eigenvectors are for instance
$|\phi\rangle \otimes |\phi\rangle$, $|\psi^1\rangle \otimes |\psi^1\rangle$ and  
$|\chi\rangle \otimes |\chi\rangle$, respectively. The functions $S_i$ are given by $S_1 =\frac{x_2^+ - x_1^-}{x_2^- - x_1^+}$, $S_2 = 1$ and $S_3 = e^{i(p_1 - p_2)} S_1$. The projectors read

\begin{eqnarray}
P_i = \frac{(C_{12} - \lambda_j)(C_{12} - \lambda_k)}{(\lambda_i - \lambda_j)(\lambda_i - \lambda_k)},
\end{eqnarray}
with $(i,j,k)$ equal to $(1,2,3)$, $(2,1,3)$ and $(3,1,2)$ respectively.  

\section{Classical $r$-matrix $r_\alg{su(1|2)}$}
After setting \cite{AF}

\begin{equation}
\label{AF}
x^\pm (x) = \frac{1}{2 \zeta} \bigg( x\sqrt{1 - \frac{\zeta^2}{(x - \frac{1}{x})^2}} \pm i \zeta \frac{x}{(x - \frac{1}{x})}\bigg), 
\end{equation}
we take $\zeta = 2 \pi / \sqrt{\lambda}$ as a deformation parameter, in the sense that we expand all formulas around $\zeta = 0$ keeping $x$ fixed. 
Since $\sin{\frac{p}{2}} = \zeta \frac{1}{x - \frac{1}{x}}$, this limit corresponds to $p \sim \zeta \sim \lambda^{-1/2}$, namely the BMN limit \cite{BMN}\footnote{We thank J. Plefka for discussion about this point.}. We also impose
$\alpha = \frac{\tilde{\alpha}}{2 \zeta}$ and $\beta = \frac{1}{2 \tilde{\alpha} \zeta}$, with $\tilde{\alpha}$ a free parameter.
The $R$-matrix and generators admit an expansion

\begin{eqnarray} 
\label{expans}
&&R_\alg{su(1|2)} \sim 1 + \zeta \, r_\alg{su(1|2)},\nonumber\\
&&\alg{J} \sim \alg{J}^{(0)} + \zeta \, \alg{J}^{(1)},\nonumber\\
&&\alg{D}_\pm \sim \alg{D}_\pm^{(0)} + \zeta \, \alg{D}_\pm^{(1)},\nonumber\\
&&e^{i p} \sim 1 + \zeta \, \pi^{(1)},
\end{eqnarray}
where the leading order generators are simply obtained using as parameters $\text{a}^{(0)},\text{b}^{(0)},\text{c}^{(0)},\text{d}^{(0)} = \lim_{\zeta \to 0} \text{a},\text{b},\text{c},\text{d}$. One has $\text{a}^{(0)}=1$, $\text{b}^{(0)} = - i \tilde{\alpha} \frac{x}{x^2 - 1}$, $\text{c}^{(0)} = \frac{i}{\tilde{\alpha} x}$, $\text{d}^{(0)} = \frac{x^2}{x^2 - 1}$. They solve a quadratic instead of a quartic constraint, namely 

\begin{equation}
\beta \, \text{a}^{(0)} \text{b}^{(0)} + \alpha \, \text{c}^{(0)} \text{d}^{(0)} = 0.
\end{equation} 
They still satisfy $\text{a}^{(0)} \text{d}^{(0)} - \text{b}^{(0)} \text{c}^{(0)} =1$, and then $\alg{J}^{(0)}$ form a $(2|2)$ representation of centrally extended $\alg{su(2|2)}$. 

If we plug the expansion (\ref{expans}) into the original formulas (\ref{coprod1}), (\ref{coprod2}) and 
(\ref{coprrel}), calling $\Delta^{triv} (x) = x \otimes 1 + 1 \otimes x$, we obtain

\begin{eqnarray}
\label{bialg}
&&[\Delta^{triv} (\alg{J}^{(0)}),r] = 0,\nonumber\\
&&[\Delta^{triv} (\alg{D}_\pm^{(0)}),r] = \pm (\alg{D}_\pm^{(0)}\otimes \pi^{(1)} -  \pi^{(1)} \otimes \alg{D}_\pm^{(0)}),
\end{eqnarray}
which means that the classical $r$-matrix is still $\alg{su(1|2)}$ invariant\footnote{One can check that (\ref{bialg}) is consistent when applied to the central charges $\alg{P}^{(0)} = \text{a}^{(0)} \text{b}^{(0)} 1$ and $\alg{K}^{(0)} = \text{c}^{(0)} \text{d}^{(0)} 1$, upon using $\pi^{(1)} = 2ix/ (x^2 - 1)$.}. One can therefore express it as a combination of projectors 

\begin{equation}
\label{exp}
r_\alg{su(1|2)} = \sum_{i=1}^3 \sigma_i P_i^{(0)},
\end{equation} 
with $P_i^{(0)} = \lim_{\zeta \to 0} P_i$, 
and with coefficient functions $\sigma_i$ given by the first order in $\zeta$ of the $S_i$, $S_i \sim 1 + \zeta \sigma_i$. One has in fact

\begin{equation}
R \sim \sum_{i=1}^3 (1 + \zeta \sigma_i )(P_i^{(0)} + \zeta P_i^{(1)}) \sim \sum_{i=1}^3 ( P_i^{(0)} + \zeta \, \sigma_i P_i^{(0)} + \zeta \, P_i^{(1)}),
\end{equation}
but since $\sum_{i=1}^3 P_i = 1$ it follows that $\sum_{i=1}^3 P_i^{(0)} = 1$ and $\sum_{i=1}^3 P_i^{(1)} = 0$. 
The quantities in (\ref{exp}) read

\begin{eqnarray}
\label{oneobt}
&&\sigma_1 = \frac{2 i (x_1^2 + x_2^2 - 2 \, x_1^2 x_2^2)}{(x_1 - x_2)(x_1^2 - 1)(x_2^2 - 1)},\nonumber\\
&&\sigma_2 = 0,\nonumber\\
&&\sigma_3 = - \frac{2 i \, x_1 x_2 (x_1^2 + x_2^2 - 2)}{(x_1 - x_2)(x_1^2 - 1)(x_2^2 - 1)},\nonumber\\
&&P_i^{(0)} = \frac{(C_{12}^{(0)} - \lambda_j^{(0)})(C_{12}^{(0)} - \lambda_k^{(0)})}{(\lambda_i^{(0)}- \lambda_j^{(0)})(\lambda_i^{(0)} - \lambda_k^{(0)})},
\end{eqnarray}
where from (\ref{proj}) one obtains in the limit $\zeta \to 0$
\begin{eqnarray}
&&\lambda_1^{(0)} = \frac{x_1^2 x_2^2}{(x_1^2 - 1)(x_2^2 - 1)},\nonumber\\
&&\lambda_2^{(0)} = \frac{x_1^2 + x_2^2}{2 (x_1^2 - 1)(x_2^2 - 1)},\nonumber\\
&&\lambda_3^{(0)} = \frac{1}{(x_1^2 - 1)(x_2^2 - 1)},
\end{eqnarray}
with $(i,j,k)$ equal to $(1,2,3)$, $(2,1,3)$ and $(3,1,2)$ respectively. 
One realizes that the $16 \times 16$ matrix $r$ projects onto a subspace of the full representation space corresponding to the eigenvalues $\lambda_1$ and $\lambda_3$ of the $\alg{su(1|2)} \otimes \alg{su(1|2)}$ Casimir. 

In the Appendix we report all entries of the classical $r$-matrix $r_\alg{su(1|2)}$. We have checked that it satisfies the standard classical (super) 
Yang-Baxter equation

\begin{eqnarray}
&&r^{i_1 i_3}_{l_1 k_3}(x_1,x_3) r^{i_2 k_3}_{l_2 l_3}(x_2,x_3) (-1)^{i_2 (i_3 + k_3)} + r^{i_1 i_2}_{l_1 j_2}(x_1,x_2) r^{j_2 i_3}_{l_2 l_3}(x_2,x_3) \nonumber\\
&&+ r^{i_1 i_2}_{j_1 l_2}(x_1,x_2) r^{j_1 i_3}_{l_1 l_3}(x_1,x_3) (-1)^{l_2 (i_3 + l_3)}  \, = \, r^{i_1 i_3}_{k_1 l_3}(x_1,x_3) r^{k_1 i_2}_{l_1 l_2}(x_1,x_2) (-1)^{i_2 (i_3 + l_3)} \nonumber\\
&&+ r^{i_2 i_3}_{j_2 l_3}(x_2,x_3) r^{i_1 j_2}_{l_1 l_2}(x_1,x_2) + r^{i_2 i_3}_{l_2 j_3}(x_2,x_3) r^{i_1 j_3}_{l_1 l_3}(x_1,x_3) (-1)^{l_2 (l_3 + j_3)}. 
\end{eqnarray}
One can see the appearance of poles in the entries: the poles at $x_i^2 = 1$ are related to singularities in the parametrization (\ref{AF}) used. The pole at $x_1 = x_2$ is present in the coefficient functions $\sigma_i$, and comes from the denominator $\frac{1}{x_2^- - x_1^+}$ in the quantum S-matrix. The residue at this pole is reported in the Appendix. There we show how to change variables 
$x_i = x_i (y_i)$
in order to make the coefficient of the residue becoming a constant matrix, and from there we read the form of the residue. It corresponds to the Casimir of the $\alg{gl(2|2)}$
superalgebra, shifted by the identity, namely in the vicinity of the pole
one has

\begin{equation}
r \sim \frac{1 \otimes 1 + \sum_{i,j=1}^4 (-)^{d[j]} E_{ij} \otimes E_{ji}}{y_1 - y_2}, 
\end{equation}
where $E_{ij}$ are the $(2|2)$ matrices with all zeroes 
up to a $1$ in the entry $(i,j)$.
The presence of the Casimir of a non-simple Lie superalgebra casts doubts on
the viability of a Belavin-Drinfeld type of analysis in the present case, but
it might nevertheless
open a way to a better understanding of the structure of the 
classical $r$-matrix and its bialgebra.

The pole at $x_1 x_2 =1$ is not present in the coefficient functions $\sigma_i$, and it corresponds to a degeneracy of the projectors when the two eigenvalues $\lambda_1^{(0)}$ and $\lambda_3^{(0)}$ coincide, as can be seen from (\ref{oneobt}). When taking into account the small $\zeta$ limit of $\sin{\frac{p}{2}} = \zeta \frac{1}{x - \frac{1}{x}}$, one can notice that this pole correspond to small momenta $p_1 + p_2 = 0$. We remind once again our interest here in the purely algebraic features of the classical $r$-matrix, while any physical interpretation should appropriately take into account the presence of the dressing factor. 

\section{Conclusions}
In this note we have obtained a solution to the standard quantum and classical 
Yang-Baxter equation from Beisert's dynamic $\alg{su(2|2)}$ spin-chain 
$S$-matrix \cite{Beisert}, 
using the manifest $\alg{su(1|2)}$-invariant formulation \cite{Beisert,Janik}.
By completely reabsorbing the lenght-changing operators action 
consistently into the bialgebra structure, 
this formulation
becomes more suitable for the Hopf-algebraic description setup in \cite{noi},
and allows a projector decomposition which simplifies the analysis of its 
properties, and
may provide insights for the reconstruction of its full 
symmetry.
We have derived the classical $r$-matrix in the parametrization
of \cite{AF}, together with its bialgebra structure, and studied
its dependence on the spectral parameters, 
especially its residue at the poles. The idea is to provide a basis for 
extracting the relevant information about the suspected enlarged
algebraic structure
underlying the integrability of the problem, 
since this is traditionally classified upon looking 
at properties of the classical $r$-matrix. The appearance of 
the Casimir of the non-simple algebra $\alg{gl(2|2)}$ in the residue at 
the simple pole at the origin, 
and the fact that the difference-form is lacking, 
makes the application of standard theorems a priori
more problematic, and represents an interesting open problem which could give 
rise to new structures, which we 
plan to investigate in the future. Developments 
in this directions appeared very recently in \cite{BeisYang}. 

An 
important step would be to make direct contact with the computations
performed from the 
string theory perspective \cite{Zaretal,FZ}. The $\alg{su(1|2)}$ decomposition
is less appealing there, therefore one might have to somehow ``covariantize'' 
the outcome of the
analysis of the enlarged symmetry algebra performed along the lines we showed  
here, before a comparison with string theory will be made.
On the other hand, the construction of the Manin triple is possibly
quite general, and one may try to adapt the procedure to the present case.
Even though we have dropped the 
phase factor as inessential to our analysis, the ultimate hope is that
the emerging symmetry, equipped with a suitable generalization of the 
crossing symmetry, will be able to determine some of its remarkable 
properties. 

\section{Acknowledgments}
The author would like to thank his collaborators, Jan Plefka and Fabian Spill,
for discussions, valuable suggestions and constant monitoring of this work, 
for careful reading of the manuscript, and for inviting him to the Physics 
Department of the Humboldt University in 
Berlin where some of these computations were finalized. We would like 
to thank Maxim Zabzine for discussions which stimulated the 
interest in studying the classical $r$-matrix, Pavel Etingof for 
useful advise 
and interesting discussions, Gleb Arutyunov, Sergey Frolov, Romuald Janik and 
Gizem Karaali
 for 
helpful email exchange, Joe Minahan for many useful discussions 
and for reading a preliminary version of the manuscript, Thomas 
Klose for useful explanations, and Daniel Grumiller for helpful 
discussions. 
We would also like to thank Antonio Bassetto and Giancarlo De Pol 
for discussions.
This work is supported in part by funds provided by the U.S. 
Department of Energy (D.O.E.) under cooperative research agreement
DE-FG02-05ER41360. The author thanks Istituto Nazionale di Fisica Nucleare
(I.N.F.N.) for supporting him through a ``Bruno Rossi'' postdoctoral 
fellowship.

\section{Appendix: The $R$-matrix $R_\alg{su(1|2)}$}

We report here the non-zero entries of the $R$-matrix $R_\alg{su(1|2)}$ (index $1$ corresponds to state $|\phi\rangle$, $2$ to $|\chi\rangle$, $3$ to $|\psi^1\rangle$ and $4$ to $|\psi^2\rangle$). As an example we can derive $R^{22}_{22}$ from Beisert's $S$-matrix:
\begin{equation}
S|\chi_1 \chi_2\rangle\ = S|\phi^2_1 Z^+ \phi^2_2 Z^+\rangle\ = e^{- i p_2} S|\phi^2_1 \phi^2_2\rangle\ = e^{- i p_2} A_{12}|\phi^2_2 \phi^2_1\rangle\ = e^{- i p_2 + i p_1} A_{12}|\chi_2 \chi_1\rangle,
\end{equation}  
where $|x_1 y_2\rangle = |x_1\rangle \otimes |y_2\rangle$.
From $R = P S$ one has then $P S |v_1^A v_2^B\rangle = P S^{AB}_{CD} (x_1,x_2) |v_2^C v_1^D\rangle = (-1)^{CD} S^{AB}_{DC} (x_1,x_2) |v_1^C v_2^D\rangle = R^{AB}_{CD} (x_1,x_2) |v_1^C v_2^D\rangle$.
One finds
\begin{eqnarray}
&&R^{11}_{11} = A_{12},\nonumber\\
&&R^{12}_{21} = \frac{1}{2} (A_{12} + B_{12}),\nonumber\\
&&R^{12}_{12} = \frac{1}{2} (A_{12} - B_{12}) e^{i p_1},\nonumber\\
&&R^{12}_{43} = -\frac{1}{2} C_{12},\nonumber\\
&&R^{12}_{34} = \frac{1}{2} C_{12},\nonumber\\
&&R^{21}_{12} = \frac{1}{2} (A_{12} + B_{12}) e^{i(p_1 - p_2)},\nonumber\\
&&R^{21}_{21} = \frac{1}{2} (A_{12} - B_{12}) e^{- i p_2},\nonumber\\
&&R^{21}_{34} = -\frac{1}{2} C_{12} e^{- i p_2},\nonumber\\
&&R^{21}_{43} = \frac{1}{2} C_{12} e^{- i p_2},\nonumber\\
&&R^{22}_{22} = A_{12} e^{i(p_1 - p_2)},\nonumber\\
&&R^{33}_{33} = - D_{12},\nonumber\\
&&R^{34}_{43} = - \frac{1}{2} (D_{12} + E_{12}),\nonumber\\
&&R^{34}_{34} = - \frac{1}{2} (D_{12} - E_{12}),\nonumber\\
&&R^{34}_{21} = \frac{1}{2} F_{12},\nonumber\\
&&R^{34}_{12} = - \frac{1}{2} F_{12} e^{i p_1},\nonumber\\
&&R^{43}_{34} = - \frac{1}{2} (D_{12} + E_{12}),\nonumber\\
&&R^{43}_{43} = - \frac{1}{2} (D_{12} - E_{12}),\nonumber\\
&&R^{43}_{21} = - \frac{1}{2} F_{12},\nonumber\\
&&R^{43}_{12} = \frac{1}{2} F_{12} e^{i p_1},\nonumber\\
&&R^{44}_{44} = - D_{12},\nonumber
\end{eqnarray}
\begin{eqnarray}
&&R^{13}_{13} = G_{12},\nonumber\\
&&R^{13}_{31} = H_{12},\nonumber\\
&&R^{14}_{14} = G_{12},\nonumber\\
&&R^{14}_{41} = H_{12},\nonumber\\
&&R^{23}_{23} = G_{12} e^{- i p_2},\nonumber\\
&&R^{23}_{32} = H_{12} e^{i(p_1 - p_2)},\nonumber\\
&&R^{24}_{24} = G_{12} e^{- i p_2},\nonumber\\
&&R^{24}_{42} = H_{12} e^{i(p_1 - p_2)},\nonumber\\
&&R^{31}_{13} = K_{12},\nonumber\\
&&R^{31}_{31} = L_{12},\nonumber\\
&&R^{41}_{14} = K_{12},\nonumber\\
&&R^{41}_{41} = L_{12},\nonumber\\
&&R^{32}_{23} = K_{12},\nonumber\\
&&R^{32}_{32} = L_{12} e^{i p_1},\nonumber\\
&&R^{42}_{24} = K_{12},\nonumber\\
&&R^{42}_{42} = L_{12} e^{i p_1}.
\end{eqnarray}
The functions appearing are Beisert's one, which we rewrite here for 
convenience after using the constraint \cite{Beisert}

\begin{eqnarray}
&&A_{12} = \frac{x_2^+ - x_1^-}{x_2^- - x_1^+},\nonumber\\
&&G_{12} = \frac{x_2^+ - x_1^+}{x_2^- - x_1^+},\nonumber\\
&&H_{12} = \frac{x_2^+ - x_2^-}{x_2^- - x_1^+},\nonumber\\
&&K_{12} = \frac{x_1^+ - x_1^-}{x_2^- - x_1^+},\nonumber\\
&&L_{12} = \frac{x_2^- - x_1^-}{x_2^- - x_1^+},\nonumber\\
&&D_{12}= - 1\nonumber\\
&&B_{12} = - 1 + \frac{(x_2^+ - x_2^- - x_1^+ + x_1^-)(x_1^+ x_2^+ - 2 x_1^- x_2^+ + x_1^- x_2^-)}{(x_2^- - x_1^+)(x_1^+ x_2^+ - x_1^- x_2^-)},\nonumber\\
&&E_{12} = A_{12} - \frac{(x_2^+ - x_2^- - x_1^+ + x_1^-)(x_1^+ x_2^+ - 2 x_2^- x_1^+ + x_1^- x_2^-)}{(x_2^- - x_1^+)(x_1^+ x_2^+ - x_1^- x_2^-)},\nonumber\\
&&C_{12} = - \frac{2 x_1^+ x_2^+ (x_2^+ - x_2^- - x_1^+ + x_1^-)}{\alpha (x_2^- - x_1^+)(x_1^+ x_2^+ - x_1^- x_2^-)},\nonumber\\
&&F_{12} = \frac{4\alpha x_1^- x_2^- (x_1^+ - x_1^-)(x_2^+ - x_2^-)(x_2^+ - x_2^- - x_1^+ + x_1^-)}{g^2 (x_2^- - x_1^+)(x_1^+ x_2^+ - x_1^- x_2^-)},\nonumber\\
&&e^{i p_i} = \frac{x_i^+}{x_i^-}.
\end{eqnarray}

\section{Appendix: The $r$-matrix $r_\alg{su(1|2)}$}
We use the parametrization in \cite{AF}:

\begin{equation}
x^\pm (x) = \frac{1}{2 \zeta} \bigg( x\sqrt{1 - \frac{\zeta^2}{(x - \frac{1}{x})^2}} \pm i \zeta \frac{x}{(x - \frac{1}{x})}\bigg) 
\end{equation}
where $\zeta = 2 \pi / \sqrt{\lambda}$ and $g^2 = \frac{1}{2 \zeta^2}$. We set as in the text
$\alpha = \frac{\tilde{\alpha}}{2 \zeta}$ and $\beta = \frac{1}{2 \tilde{\alpha} \zeta}$, with $\tilde{\alpha}$ a free parameter. 

The non-zero entries of the classical $r$-matrix $r_\alg{su(1|2)}$ read

\begin{eqnarray}
&&r^{11}_{11} = \frac{2 i (x_1^2 + x_2^2 - 2 x_1^2 x_2^2)}{(x_1 - x_2)(x_1^2 - 1)(x_2^2 - 1)},\nonumber\\
&&r^{12}_{21} = - \frac{2 i x_1 x_2}{(x_1 - x_2)(x_1 x_2 - 1)},\nonumber\\
&&r^{12}_{12} = - \frac{2 i x_2 (x_1^3 + x_2 - x_1 x_2^2 - 2 x_1^2 x_2 + x_1^2 x_2^3)}{(x_1 - x_2)(x_1^2 - 1)(x_2^2 - 1)(x_1 x_2 - 1)},\nonumber\\
&&r^{12}_{43} = - r^{12}_{34} = -\frac{2}{\tilde{\alpha} (x_1 x_2 - 1)},\nonumber\\
&&r^{22}_{22} = - \frac{2 i x_1 x_2 (x_1^2 + x_2^2 - 2)}{(x_1 - x_2)(x_1^2 - 1)(x_2^2 - 1)},\nonumber\\
&&r^{33}_{33} = r^{44}_{44} = 0,\nonumber\\
&&r^{34}_{43} = - r^{34}_{34} = \frac{2 i x_1 x_2}{(x_1 - x_2)(x_1 x_2 - 1)},\nonumber\\
&&r^{34}_{21} = - r^{34}_{12} = \frac{2 \tilde{\alpha} x_1^2 x_2^2}{(x_1^2 - 1)(x_2^2 - 1)(x_1 x_2 - 1)},\nonumber\\
&&r^{13}_{13} = r^{13}_{31} = r^{14}_{14} = r^{14}_{41} = \frac{2 i x_2^2}{(x_1 - x_2)(1 - x_2^2)},\nonumber\\
&&r^{23}_{23} = r^{24}_{24} = \frac{2 i x_1 x_2}{(x_1 - x_2)(1 - x_2^2)},\nonumber\\
&&r^{23}_{32} = r^{24}_{42} = \frac{2 i x_2^2}{(x_1 - x_2)(1 - x_2^2)},
\end{eqnarray}
and one can obtain the other ones not displayed by using unitarity, namely

\begin{equation}
r^{AB}_{CD} (x_2,x_1) + (-)^{AB + CD} r^{BA}_{DC} (x_1,x_2) = 0,
\end{equation}
which ensues from unitarity of the quantum $S$-matrix $S_{12} S_{21} = 1$.

The residue at the pole $x_1 = x_2$ reads

\begin{eqnarray}
\label{resid}
&&r^{11}_{11} = r^{22}_{22} = \frac{- 4 i x_1^2 }{x_1^2 - 1},\nonumber\\
&&r^{12}_{21} = r^{12}_{12} = r^{34}_{34} = - r^{34}_{43} = \frac{- 2 i x_1^2 }{x_1^2 - 1},\nonumber\\
&&r^{12}_{43} = r^{12}_{34} = r^{33}_{33} = r^{34}_{21} = r^{34}_{12} = r^{44}_{44} = 0,\nonumber\\
&&r^{13}_{13} = r^{13}_{31} = r^{14}_{14} = r^{14}_{41} = \frac{- 2 i x_1^2 }{x_1^2 - 1},\nonumber\\
&&r^{23}_{23} = r^{23}_{32} = r^{24}_{24} = r^{24}_{42}= \frac{- 2 i x_1^2 }{x_1^2 - 1}
\end{eqnarray}
and using again unitarity one can obtain the not displayed entries. In 
(\ref{resid}), $r^{ij}_{kl}$ is an abuse of notation for the residue
$Res[r^{ij}_{kl}, x_1 = x_2]$).

In terms of the projector decomposition (\ref{exp}), 
the residue is easily computed as $\frac{- 4 i x_1^2 }{x_1^2 - 1} 
[{P_1^{(0)}}_1 + {P_3^{(0)}}_1]$, where ${P_i^{(0)}}_1$ project onto
irreducible representations of $\alg{su(1|2)} \otimes \alg{su(1|2)}$ where 
the same representation $1$ is chosen on both sides of the tensor product.
This can be rewritten as

\begin{equation}
\label{resappend}
r \sim \frac{f(x_1)(1 \otimes 1 + \sum_{i,j=1}^4 (-)^{d[j]} E_{ij} \otimes E_{ji})}{x_1 - x_2}, 
\end{equation}
where $E_{ij}$ are the $(2|2)$ coordinate-matrices with all zeroes 
up to a $1$ in the entry $(i,j)$, and $f(x_1) = - \frac{2i x_1^2}{x_1^2 - 1}$.

There is a standard way of changing variables in order to reduce the residue
at the pole to a constant matrix \cite{BDchange}. We notice that this trick 
works in the simple Lie algebra case, and for $r$-matrices of difference-form, 
therefore it is not a priori guaranteed that
one can apply it in the present case as well. This is due to the particular 
form of the residue (\ref{resappend}), 
and it may signify some important features of this 
$r$-matrix.

Suppose the behaviour in the vicinity of the pole is 
$r \sim \frac{f(x_1) t}{x_1 - x_2}$, with $t$ a constant matrix, and 
$f(x_1)\neq0$. Then, one can easily verify that 
changing variables $x = x(y)$ in such a way that $x'(y) = f(x(y))$,
allows to reduce the function $f$ to $1$. In our case we have the following 
differential equation:

\begin{equation}
x'(y) = - \frac{2i x(y)^2}{x(y)^2 - 1},
\end{equation}
which is solved by 
$x(y) = \frac{1}{2} (- 2i y + c \pm \sqrt{c^2 - 4i c y - 4 y^2 - 4})$, 
$c$ being an integration constant.

\end{document}